\documentclass[a4paper,12pt]{article}
\usepackage{amssymb}
\usepackage{graphicx}
\usepackage{cite}

\makeatletter
\def\opt@space{\@ifnextchar,\relax{\@ifnextchar\ \relax{\@ifnextchar:\relax{\@ifnextchar.\relax { }}}}}
\let\optspace=\opt@space
\makeatother

\DeclareRobustCommand{\ie}{{\it i.e.}\optspace}
\DeclareRobustCommand{\eg}{{\it e.g.}\optspace}

\def\ap#1#2#3   {{\em Ann. Phys. (NY)} {\bf#1} (#2) #3.}   
\def\apj#1#2#3  {{\em Astrophys. J.} {\bf#1} (#2) #3.} 
\def\apjl#1#2#3 {{\em Astrophys. J. Lett.} {\bf#1} (#2) #3.}
\def\app#1#2#3  {{\em Acta. Phys. Pol.} {\bf#1} (#2) #3.}
\def\ar#1#2#3   {{\em Ann. Rev. Nucl. Part. Sci.} {\bf#1} (#2) #3.}
\def\cpc#1#2#3  {{\em Computer Phys. Comm.} {\bf#1} (#2) #3}
\def\err#1#2#3  {{\it Erratum} {\bf#1} (#2) #3.}
\def\epj#1#2#3  {{\it Eur. Phys. J.} {\bf#1} (#2) #3}
\def\ib#1#2#3   {{\it ibid.} {\bf#1} (#2) #3.}
\def\jmp#1#2#3  {{\em J. Math. Phys.} {\bf#1} (#2) #3.}
\def\ijmp#1#2#3 {{\em Int. J. Mod. Phys.} {\bf#1} (#2) #3}
\def\jetp#1#2#3 {{\em JETP Lett.} {\bf#1} (#2) #3.}
\def\jpg#1#2#3  {{\em J. Phys. G.} {\bf#1} (#2) #3}
\def\mpl#1#2#3  {{\em Mod. Phys. Lett.} {\bf#1} (#2) #3.}
\def\nat#1#2#3  {{\em Nature (London)} {\bf#1} (#2) #3.}
\def\nc#1#2#3   {{\em Nuovo Cim.} {\bf#1} (#2) #3.}
\def\nim#1#2#3  {{\em Nucl. Instr. Meth.} {\bf#1} (#2) #3.}
\def\np#1#2#3   {{\em Nucl. Phys.} {\bf#1} (#2) #3}
\def\pcps#1#2#3 {{\em Proc. Cam. Phil. Soc.} {\bf#1} (#2) #3.}
\def\pl#1#2#3   {{\em Phys. Lett.} {\bf#1} (#2) #3}
\def\prep#1#2#3 {{\em Phys. Rep.} {\bf#1} (#2) #3}
\def\prev#1#2#3 {{\em Phys. Rev.} {\bf#1} (#2) #3}
\def\prl#1#2#3  {{\em Phys. Rev. Lett.} {\bf#1} (#2) #3}
\def\prs#1#2#3  {{\em Proc. Roy. Soc.} {\bf#1} (#2) #3.}
\def\ptp#1#2#3  {{\em Prog. Th. Phys.} {\bf#1} (#2) #3.}
\def\ps#1#2#3   {{\em Physica Scripta} {\bf#1} (#2) #3.}
\def\rmp#1#2#3  {{\em Rev. Mod. Phys.} {\bf#1} (#2) #3}
\def\rpp#1#2#3  {{\em Rep. Prog. Phys.} {\bf#1} (#2) #3.}
\def\sjnp#1#2#3 {{\em Sov. J. Nucl. Phys.} {\bf#1} (#2) #3}
\def\spu#1#2#3  {{\em Sov. Phys.-Usp.} {\bf#1} (#2) #3.}
\def\zp#1#2#3   {{\em Zeit. Phys.} {\bf#1} (#2) #3}

\newcommand\Fig[1]{Fig.~\ref{#1}}

\newcommand\Eq[1]{(\ref{#1})}

\DeclareRobustCommand\ET{\ensuremath{E_{\rm T}}\optspace}

\DeclareRobustCommand\Fg{\ensuremath{F_2^\gamma}\optspace}
\DeclareRobustCommand\Nf{{\ensuremath{N_{\rm f}}}\optspace}
\newcommand{\sclev}{{\cal S}_{\rm ev}}
\newcommand{\sclps}{{\cal S}_{\rm PS}}
\newcommand{\xh}{\hat x}
\newcommand{\as}[1]{\ensuremath{\alpha_{\rm s}(#1^2)}\optspace}
\newcommand{\asop}{\ensuremath{{\alpha_{\rm s} \over 2 \pi}}\optspace}
\newcommand{\aop}[1]{\ensuremath{{\alpha_{\rm s}(#1^2) \over 2 \pi}}\optspace}
\newcommand{\LQCD}{\ensuremath{\Lambda_{\rm QCD}}\optspace}
\newcommand\qb{\bar q}
\newcommand{\GeV}{{\rm GeV}}
\newcommand{\aem}{{\ensuremath{\alpha_{\rm em}}\optspace}}
\newcommand{\PL}{{\rm PL}}
\newcommand{\HL}{{\rm HAD}}
\newcommand{\MSbar}{{\ensuremath{\overline{\rm MS}}}\optspace}
\newcommand{\DISg}{{\ensuremath{{\rm DIS}\gamma}}\optspace}

\DeclareRobustCommand{\Sigl}{{\ensuremath{\Sigma'}}\optspace}

\newcommand{\Hh}{\ensuremath{{\cal H}_h}\optspace}
\newcommand{\Has}{\ensuremath{{\cal H}^{\rm as}_h}\optspace}
\newcommand{\HG}{\ensuremath{{\cal H}^G_h}\optspace}
\newcommand{\Hgam}{\ensuremath{{\cal H}^\gamma_h}\optspace}

\begin{document}
\title{NLO photon parton parametrization using $ee$ and $ep$ data%
\footnote{Web~page:~http://th-www.if.uj.edu.pl/$\sim$wojteks/SAL}}
\author{%
{\bf W.~Slominski\footnote{e-mail address:
wojteks@th.if.uj.edu.pl}}\\
{\small \sl M. Smoluchowski Institute of Physics, Jagellonian University}\\ 
{\small \sl Reymonta 4, 30-059 Cracow, Poland}\\[3mm]
{\small and}\\[3mm]
 {\bf H.~Abramowicz\footnote{also at Max Planck Institute, Munich, Germany, Alexander von 
Humboldt Research Award.}, A.~Levy} \\ 
{\small \sl School of Physics and Astronomy,}\\
{\small \sl Raymond and 
Beverly Sackler Faculty of Exact Sciences}\\
  {\small \sl Tel--Aviv University, Tel--Aviv, Israel}
}
\date{}
\maketitle

\begin{abstract}
  An NLO photon parton parametrization is presented based on the
  existing $F_2^\gamma$ measurements from $e^+e^-$ data and the
  low-$x$ proton structure function from $ep$ interactions. Also
  included in the extraction of the NLO parton distribution functions
  are the dijets data coming from $\gamma p \to j_1 + j_2 +X$. The
  new parametrization is compared to other NLO
  parametrizations.
\end{abstract}

\setcounter{page}{0}
\thispagestyle{empty}
\newpage  

\section{Introduction}

In spite of the photon being a fundamental gauge particle of
electromagnetic interactions, it is known to develop a hadronic
structure in its interactions with matter.  The notion of the hadronic
structure function of the photon, $F_2^{\gamma}$, is introduced in
analogy to the well known nucleon case. The first measurements of \Fg
became available from $e^+ e^-$ collisions in which one of the leptons
is scattered under a small angle. These interactions may then be
interpreted as processes in which a highly virtual photon, of
virtuality $Q^2$, probes an almost real target photon, with virtuality
$P^2 \approx 0$.

While the proton structure function $F_2^p$ has been well measured
over a wide range of $Q^2$ and the Bjorken variable
$x$~\cite{reviewf2}, $F_2^{\gamma}$ data cover a restricted kinematic
range ($0.001 < x < 0.9$) and are subject to much larger systematic
uncertainties~\cite{hawar}. This is mainly due to experimental
limitations in measuring the centre of mass energy $W$ of the
$\gamma^*\gamma$ system, in particular at large values of $W$.

In the present paper, a new parametrization of the parton
distributions in the photon is extracted in next-to-leading order
(NLO) of perturbative QCD. It differs from other NLO
parametrizations~\cite{GRV92,AFG,GS96,GRS99,CJK} in that the data used
in the fitting procedure include the expected behaviour of \Fg at
low-$x$~\cite{al-gribov}, as derived from $F_2^p$
measurements~\cite{f2p} under Gribov factorization
assumption~\cite{gribov-fact} and, in addition, the measurements of
the dijet photoproduction cross sections~\cite{dijets}.

\section{Theoretical background}

\subsection{Physical quantities and parton distributions}

In QCD, the hadronic cross sections are given as convolutions of
coefficient functions and parton densities.  In higher orders of
perturbative expansion, the definition of parton densities is not
unique and depends on the adopted factorization scheme.
In this paper we adopt
the \DISg factorization scheme~\cite{GRV92}, where the relation between
\Fg and the parton distributions,
$f^\gamma_k(x,Q^2)$ ($k=q \;\rm{or}\; G$), is given by
\begin{eqnarray}
\label{eq:F2g}
\bar\Fg(Q^2) &=&
\left[ 1 + \aop{Q}\, C_{F,2}^{(1)} \right]
\otimes
\sum_{q=1}^{\Nf} 2 e_q^2 f^\gamma_{q}(Q^2)
\nonumber\\
&&+ \left( \sum_{q=1}^{\Nf} e_q^2 \right)
\, \aop{Q}\, 
C_{G,2}^{(1)} \otimes f^\gamma_{G}(Q^2)
\,,
\end{eqnarray}
where
$\bar\Fg(x,Q^2) \equiv \frac{1}{x}\Fg(x,Q^2)$, 
$\as{Q}$ is the QCD running coupling constant,
$\otimes$ denotes convolution,
\Nf is the number of flavours and $e_q$ is the electric charge of
quark $q$ in
units of $e$. The parton distributions are convoluted with known
coefficient functions%
\footnote{$C_{G,2}^{(1)}$ used here is the
one given in Appendix I of \cite{FP} divided by \Nf.},
 $C_{F,2}(x),C_{G,2}(x)$~\cite{FP,GRV92}. 
In Eq.~\Eq{eq:F2g}, all the \Nf flavours are assumed to be massless
and the effects of heavy quarks are discussed in Sec.~\ref{s:heavy}.

The cross-section for dijets photoproduction depends on both the
photon and the proton parton distributions and we take the latter to
be fixed by the deep inelastic $ep$ scattering measurements. To
calculate this cross-section at NLO, the program by Frixione and
Ridolfi~\cite{Frix} 
is used.

\subsection{Evolution equations for the photon}

In order to model the parton content of the photon,
$f^\gamma_k\!\left(x,Q^2\right)$ ($k = q,\qb,G$), we parametrize the
parton distributions (PDF) at some scale $Q_0$,
$f^\gamma_k\!\left(x,Q_0^2\right)$, and evolve them to other scales 
through the analogue of the DGLAP evolution equations for the photon 
which contain an inhomogeneous term~\cite{witten,bardeen,dewitt}.
\begin{equation}
\label{eq:evol}
{df^\gamma_k\!\left(x,Q^2\right) \over d\ln Q^2}
=
{\aem \over 2 \pi} P_{k\gamma}\!\left(x,Q^2\right)
+ \aop{Q} 
\int\limits_x^1\! {d\xh \over \xh}
\sum_{{j=\atop q,\qb,G}} P_{kj}\!\left({x\over\xh},Q^2\right) 
f^\gamma_j\!\left(\xh,\ Q^2\right) 
\,,
\end{equation}
where
$P_{kA}\left(x,Q^2\right)$ are the splitting functions,
and \aem is the fine-structure constant.

Within next-to-leading order accuracy the splitting functions are
given by
\begin{equation}
P_{kA}\left(x,Q^2\right) = P_{kA}^{(0)}(x) + \aop{Q} P_{kA}^{(1)}(x)
\,,
\end{equation}
and the higher order terms $P_{kA}^{(1)}(x)$ depend on the adopted
factorization scheme.  The QCD splitting functions for $A = q,\qb,G$ at
NLO in the \MSbar scheme are given in \cite{CFP}.  The photon splitting
functions, $P_{k\gamma}\left(x,Q^2\right)$ can be obtained from the
gluonic ones ($P_{qG},P_{GG}$) by setting $C_G = 0$ and multiplying by
factor three.
For the \DISg factorization scheme the latter still have to be
transformed as described in \cite{GRV92}.

\subsection{Heavy quarks}
\label{s:heavy}

In order to solve the evolution equations given by \Eq{eq:evol} we
still need to specify the treatment of the heavy flavour contribution.

The approaches traditionally used are either the Fixed Flavour Number
scheme (FFNS) or the Zero Mass Variable Flavour Number scheme
(ZM-VFNS). While FFNS is known to be adequate at low $Q^2$ values, the
ZM-VFNS is the appropriate scheme for calculations at high $Q^2$.
Recent developments \cite{ACOT,Collins,TR,Tung2002} 
aim at constructing schemes which work
well in the whole $Q^2$ range. Any such scheme has to approach FFNS at
low $Q^2$ and ZM-VFNS at high $Q^2$.  In the following we propose a
phenomenological approach which smoothly interpolates between FFNS and
ZM-VFNS results.

In general one can write
\begin{eqnarray}
\bar\Fg(Q^2) &=&
\left[ 1 + \aop{Q}\, C_{F,2}^{(1)} \right]
\otimes
\sum_{q=d,u,s} 2 e_q^2 f^\gamma_{q}(Q^2)
\nonumber\\
&+&
\left( \sum_{q=d,u,s} e_q^2 \right)\,
 \aop{Q}\, C_{G,2}^{(1)} \otimes f^\gamma_{G}(Q^2)
\nonumber\\
&+&
\sum_h e_h^2 \, \Hh(Q^2)
\,,
\end{eqnarray}
where the contributions from the light and heavy quarks, $e_h^2 \,
\Hh(Q^2)$, are given separately.

In FFNS there are no heavy quarks in the probed target and a pair of
heavy quarks can only be produced in the final state.  The threshold
condition for such production is for the hadronic state mass, $W$, to
be above $2 m_h$, \ie
\begin{equation}
{W^2} = Q^2 {1-x\over x} > 4 m_h^2
\,,
\end{equation}
where $m_h$ is the heavy quark mass.
$\Hh(Q^2)$ is calculated at fixed order in \aem and \as{Q} with $m_h
\neq 0$.  Within the accuracy needed for the current calculation only
$\gamma^*\gamma \to h\bar h$ and $\gamma^*G \to h\bar h$ processes
contribute, \ie $\Hh = \Hgam +\HG$.  \def\sighh{\hat\sigma_{hh}} The
point-like photon contribution from $\gamma^*\gamma$ scattering,
$\Hgam$, is
\begin{equation}
\Hgam(x,Q^2) = 6 e_h^2 {\aem\over 2\pi}
\,\sighh(x,Q^2)
\label{eq:hg}
\end{equation}
while the contribution from $\gamma^*G$ scattering, $\HG$, is 
\begin{equation}
\HG(x,Q^2) = {\as{Q}\over 2\pi}
\int\limits_x^1 \, {dz\over z}\,
f^\gamma_G(x/z,Q^2)
\,\sighh(z,Q^2)
\,.
\label{eq:hG}
\end{equation}
$\sighh$ is the Bethe-Heitler type reduced cross
section~\cite{Budnev,GRV92,AFG},
\goodbreak
\begin{eqnarray}
\sighh(x,Q^2) &=&
\Theta(W-2m_h)\, 
\nonumber\\&\times&
\Bigg\{
\left[
x^2 + (1-x)^2 + 4 x (1-3x) {m_h^2\over Q^2} - 8 x^2 {m_h^4\over Q^4}
\right]
\nonumber\\&&\times
\ln {\left(1+\sqrt{1-\beta}\right)^2 \over \beta}
\nonumber\\
&&+
\left[
8x(1-x) -1 - 4 x (1-x) {m_h^2\over Q^2}
\right] \sqrt{1-\beta}
\Bigg\}
\,,
\label{eq:h-BH}
\end{eqnarray}
where 
$\displaystyle\beta = {4 m_h^2 x\over Q^2 (1-x)} = {4 m_h^2\over W^2}$.

For $Q^2 \gg m_h^2$ we can neglect terms vanishing in the limit
$m_h^2/Q^2 \to 0$, which results in the approximation
\begin{eqnarray}
\sighh(x,Q^2) &\simeq&
\left[ x^2 + (1-x)^2 \right]
\ln {Q^2 \over m_h^2}
\nonumber\\
&& +
\left[ x^2 + (1-x)^2 \right]
\ln {1-x \over x}
+ 8x(1-x) - 1
\,.
\label{eq:BH-as}
\end{eqnarray}
The large logarithm, $\ln{Q^2\over m_h^2}$, spoils the accuracy of this
fixed order calculation and in order to achieve the NLO accuracy this
term would have to be resummed. This resummation would 
generate the $f^\gamma_h$ distribution.

In the ZM-VFNS the
heavy quark masses, $m_h$, serve only as transition scales. When $Q^2$
crosses the value of $m_h^2$, the number of active flavours, \Nf,
changes by one.  The effect on the evolution equations is twofold.
First, the limit of summation over quarks changes and, second, the
$Q^2$ dependence of \as{Q} changes.  As we want \as{Q} to be
continuous across the transition scales, we adjust the scale parameter
of QCD, $\LQCD(\Nf)$, so that
\begin{equation}
\alpha_{\rm s}(m_h^2; \Nf, \LQCD(\Nf)) = 
\alpha_{\rm s}(m_h^2; \Nf+1, \LQCD(\Nf+1)) 
\,.
\end{equation}
Otherwise there is no dependence on quark masses in the evolution.
The evolution is done according to Eq.~\Eq{eq:evol} and \Hh is given
by the same formula as for light quarks, 
\begin{eqnarray}
\Hh(Q^2) &=&
2 \left[ 1 + \aop{Q}\, C_{F,2}^{(1)} \right]
\otimes
f^\gamma_h(Q^2)
\nonumber\\
&& + \,\aop{Q}\, C_{G,2}^{(1)} \otimes f^\gamma_{G}(Q^2)
\nonumber\\
&\equiv&
\Has(Q^2)
\,.
\label{eq:h-as}
\end{eqnarray}
This ZM-VFNS formula sums up large collinear logarithms which were
discussed in the context of validity of \Eq{eq:BH-as}. The remaining
terms of \Eq{eq:BH-as}, contributing to the $\gamma^*\gamma$ process,
are included in the $f^\gamma_h$ definition in the \DISg scheme. The
same terms in the $\gamma^*G$ contribution lead to the second
term of \Has (see Eq.~\Eq{eq:h-as}). However, at finite $Q^2$ the
dependence on $m_h$ starts to be important and in a more precise
calculation~\cite{Collins} it is taken into account by the mass
dependent coefficient functions.

In summary, FFNS has no resummation of the large collinear logarithms,
important at high $Q^2$, and ZM-VFNS has no mass dependence, important
at low $Q^2$ and near the $W = 2m_h$ threshold.
Note that both ZM-VFNS and FFNS schemes are formally correct pQCD results.
At intermediate scales the contributions from both powers of $m_h^2/Q^2$
(Eq.~\Eq{eq:h-BH}) and collinear logs resummation (Eq.~\Eq{eq:h-as}),
should be taken into account. What we are looking for now is a
prescription which would be a good approximation over the whole range
of $Q^2$.

It is clear from the above discussion that simply adding the two
contributions would result in some double counting.  A consistent
solution is not uniquely given by QCD \cite{Thorne-DIS05} and several
approaches have been discussed~\cite{TR,ACOT,Tung2002,CJK}. Here, we
propose to construct the heavy quark contribution to \Fg as a weighted
sum of FFNS and ZM-VFNS expressions
with a $Q^2$-dependent weight, $\sclev$,
\begin{equation}
\Hh(Q^2) = \left[1-\sclev(m_h^2,Q^2)\right] (\Hgam(Q^2) + \HG(Q^2)) 
+ \sclev(m_h^2,Q^2) \Has(Q^2)
\label{Hw0}
\,.
\end{equation}
Intuitively, $\sclev$ quantifies the amount of evolution and it will
be defined so that $\sclev =0$ at $Q^2 \leq m_h^2$ and $\sclev \to 1$
for $Q^2 \gg m_h^2$. By construction, the coefficients of $(\Hgam +
\HG)$ and \Has in \Eq{Hw0} add up to 1 to prevent double counting.
Using \Eq{eq:h-as} we can explicitly write out the quark and gluon
contributions
\begin{eqnarray}
\Hh &=&
(1-\sclev) \Hgam
+ 2 \sclev
\left[ 1 + \asop\, C_{F,2}^{(1)} \right]
\otimes
f^\gamma_h
\nonumber\\
&+& (1-\sclev) \asop\, \sighh \otimes f^\gamma_{G}
+ \sclev\asop\, C_{G,2}^{(1)} \otimes f^\gamma_{G}
\,.
\end{eqnarray}
The NLO corrections coming from the $\gamma^*G \to h\bar h$ process
are given by the last two terms where $\sighh$
plays the role of the mass dependent coefficient function. As this is an
NLO contribution and we have no mass dependence in the other
coefficient functions, we take
\begin{eqnarray}
\Hh &=&
(1-\sclev) \Hgam
+ 2 \sclev
\left[ 1 + \asop\, C_{F,2}^{(1)} \right]
\otimes
f^\gamma_h
\nonumber\\
&+& \asop\, C_{G,2}^{(1)} \otimes f^\gamma_{G}
\,.
\end{eqnarray}
This last approximation has the advantage of noticeably speeding up
the minimization procedure, at a cost of disregarding a small correction,
well within the uncertainty of the NLO approximation.

We define $\sclev$ as
\begin{equation}
\sclev(m_h^2,Q^2) = \cases {
\displaystyle {0 \;\mbox{ for }\; Q^2 \leq m_h^2\hfill}
\,,
\atop
\displaystyle\min\left(1,\tau(m_h^2,Q^2)\right) \;\mbox{ for }\; Q^2 > m_h^2
\,,}
\end{equation}
where
\begin{equation}
\tau(m_h^2,Q^2) =
\ln
{\ln{Q^2\over\Lambda^2_4}
 \over
\ln{m_h^2\over\Lambda^2_4}}
\end{equation}
is the QCD evolution scale.

To correct the behaviour of massless coefficient functions close to the
$W$ threshold, we introduce a
phase space suppression factor, $\sclps$.
$\sclps$ must be zero at the threshold, $W = 2m_h$, and must go to one
for $W \gg 2m_h$.  $\sclps$ is taken as the ratio of the leading twist
\Hgam over its asymptotic ($Q^2 \gg m_h^2$) value,
\begin{equation}
\sclps(m_h^2/Q^2, x) = 
{
\left[x^2 + (1-x)^2\right] \ln{(1+\sqrt{1-\beta})^2 \over \beta}
+ \left[8x(1-x) - 1\right] \sqrt{1-\beta}
\over
\left[x^2 + (1-x)^2\right] 
\ln{4 \over \beta}
+ 8x(1-x) - 1
}\,.
\end{equation}

The final result for heavy quark contribution to \Fg is
\begin{eqnarray}
\Hh(Q^2) &=&
\left[1-\sclev(m_h^2,Q^2)\right] \Hgam(Q^2) 
\nonumber\\
&+& 2\, \sclev(m_h^2,Q^2)\, \sclps(m_h^2/Q^2)\,
\left[ 1 + \aop{Q}\, C_{F,2}^{(1)} \right]
\otimes
f^\gamma_h(Q^2)
\nonumber\\
&+& \sclps(m_h^2/Q^2)\, \aop{Q}\, C_{G,2}^{(1)} \otimes f^\gamma_{G}(Q^2)
\,.
\end{eqnarray}

In the NLO calculation of the dijet photoproduction cross sections,
as implemented in the
Frixione and Ridolfi program~\cite{Frix}, the partonic
cross sections assume massless quarks. The
factorization scale is taken as the average $\ET$ of the jets.
The value of \ET for the data considered in this paper is well above
$m_b$ and below $m_t$. Hence we expect the heavy quarks mass effects
to be negligible.

\subsection{Constraints on the gluons} \label{sec:glu}
\label{sec:F-G}
The gluon content of the photon can be extracted only indirectly, in
particular from the $Q^2$ evolution of \Fg. The case of the photon is
even more challenging than that of the proton. The contribution of the
point-like splitting of the photon into a $q\bar{q}$ is at the origin
of strong scaling violations, positive for all values of $x$. This
decreases the sensitivity to scaling violations due to gluon
radiation. The difficulty is exacerbated further by the lack of the
momentum sum rule and of small $x$ measurements where the scaling
violation is dominated by gluons.

In the present extraction, the lacking small $x$ measurements of \Fg
are replaced by the low $x$ measurements of the proton $F_2$
transformed as suggested in \cite{al-gribov}. The transformation is
based on Gribov factorization~\cite{gribov-fact} which relates the
total $\gamma\gamma$ cross section to those of $\gamma p$ and $pp$.
For low $x$ one can thus obtain
\begin{equation}
\Fg(x,Q^2)= F_2^p(x,Q^2)\frac{\sigma_{\gamma p}(W)}{\sigma_{pp}(W)}.
\end{equation}
Using the Donnachie and Landshoff~\cite{dl} (DL) parametrization of the
cross sections, which gives a good representation of the data, one
obtains at large $W$
\begin{equation}
\Fg/\aem = 0.43\, F_2^p.
\label{eq:gdl}
\end{equation}
In the following, we assume that Gribov factorization holds and only
assign an error to the transformation~(\ref{eq:gdl}) which arises from
the DL parametrization. More specifically, we refitted the $pp$,
$p\bar{p}$ and $\gamma p$ total cross section data at high energy
keeping the Regge intercepts obtained by DL.  Depending on the energy
range fitted, the uncertainty on the ratio $\sigma_{\gamma
  p}(W)/\sigma_{pp}(W)$ varied anywhere between 1\% and 6\% and we
chose to use 3\% as an error estimate.

Frankfurt and Gurvich derived an equivalent of the momentum sum rule,
the FG sum rule~\cite{fg-sumrule}\footnote{A similar sum rule in LO was
derived by \protect\cite{SaS}.}. It relates the momentum fraction
carried by partons in the resolved photon to the physical $e^+ e^-
\to$ ``hadrons'' cross section. The authors estimate that for $Q^2 <
10\, \GeV^2$
\begin{equation}
\label{eq:FG}
{1\over\aem}\,
\int\limits_0^1 \! dx\, x \left[ 
\Sigma'(x,Q^2, P^2=0) + G(x,Q^2, P^2=0)
\right]
\approx 1+
{2\over 3\pi} \ln{Q^2\over 4\,\GeV^2}
\,,
\end{equation}
where
\begin{equation}
\Sigma'(x,Q^2,0) =
\sum\limits_{q=u,d,s} \left[ f^\gamma_q(x,Q^2) + f^\gamma_{\bar q}(x,Q^2)
\right]
\end{equation}
and $G(x,Q^2, P^2=0) = f^\gamma_G(x,Q^2)$.  The quark distributions in
the photon used in \cite{fg-sumrule} are normalized such that
$\Sigma'(x,Q^2,Q^2)$ is given by the box diagram contribution. At NLO
this normalization corresponds to the \DISg factorisation scheme,
where the subleading box diagram contribution is absorbed into the
quark distributions.

This sum rule was not directly incorporated in the fits, however it is
used to assess the quality of the fitted parametrization.

\begin{figure}[ht]
\centerline{\includegraphics*[width=\columnwidth]{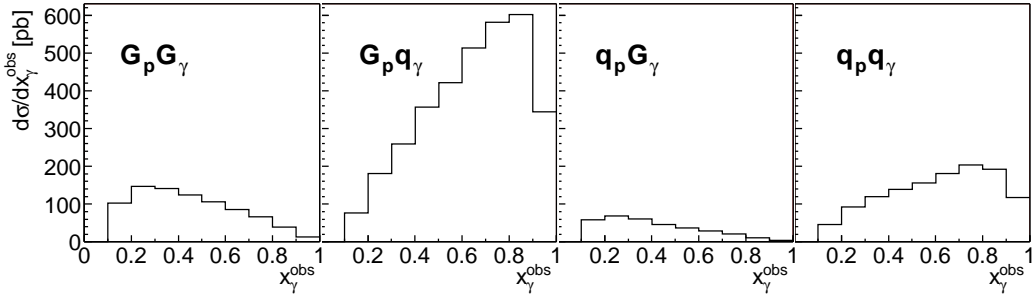}}
\caption{Typical contributions of different
parton types, as denoted in the figure, to ${d\sigma / dx_\gamma^{\rm obs}}$
for $\ET \in [14, 17]\,
\GeV$.  Hadronization corrections are not included.}
\label{fig:djpart}
\end{figure}

In an attempt to constrain further the gluon distribution in the
photon, the cross section measurements of dijet photoproduction were
added in the fitting procedure. However we found out that these
measurements are dominated by the contribution of gluons in the
proton, while the region most sensitive to gluons in the photon is
suppressed by kinematic constraints.  To establish this, we calculated
the resolved photon contributions to the total dijet cross section
from various partonic subprocesses as a function of the observed
photon momentum fraction participating in the hard interaction,
$x_{\gamma}^{\rm obs}$.  The cross sections were calculated using
the Frixione--Ridolfi code with CTEQ5M proton PDFs and our photon PDFs.
The results are shown in \Fig{fig:djpart}. 
Note that the dominant contribution depends on the quark content of
the photon which is fairly well established by the \Fg data.  We
checked that this conclusion is not affected by the choice of the PDFs
used in the calculation.
\section{Parametrization}

In practice we solve the NLO evolution equations numerically in the
$x$ space. We use a programme \cite{ws-evol} which
performs the evolution over a two-dimensional grid in $x$ and $\ln (Q^2)$.
The actual $f^\gamma_k\left(x,Q^2\right)$ values are obtained from the
grid by a quadratic interpolation.

The dijet cross sections, when included in the fit, are recalculated
for each set of $f^\gamma_k\left(x,Q^2\right)$ and the hadronic
corrections are added.

Our parametrization of the initial parton distributions, defined at
$Q^2_0= 2\,\GeV^2$, aims at describing the experimental data below the charm
threshold. Thus we explicitely parametrize only the $u,d,s$ quarks and
the gluon. The $c,b$ and $t$ quarks are generated radiatively once
their respective thresholds (transition scales) are crossed.

All quark distributions in the photon are parametrized as a sum of 
point-like and hadron-like contributions,
\begin{equation}
f_q(x) = f_{\qb}(x) =
 e_q^2 {A^\PL} {x^2+(1-x)^2 \over 1- {B^\PL} \ln(1-x)}
   + f_q^\HL(x)
\,.
\end{equation}

Apart from the $e_q^2$ factor, the point-like contribution is the same
for all quarks. The hadron-like contribution is assumed to depend on
the quark mass only. For $u$ and $d$ quarks we parametrize it as
\begin{equation}
f_u^\HL(x) = f_d^\HL(x)
 = {A^\HL} x^{{B^\HL}} (1-x)^{{C^\HL}}
\,.
\end{equation}

We fix the $s$ quark distribution to be
\begin{equation}
f_s^\HL(x) = 0.3\, f_d^\HL(x)
\,.
\end{equation}

The gluons in the photon are assumed to have hadron-like behaviour
\begin{equation}
  f_G(x) = {A^\HL_G} x^{{B^\HL_G}} (1-x)^{{C^\HL_G}}
\,.
\end{equation}

As there are no data at $x\simeq 1$ we fix 
$C^\HL = 1$ and $C^\HL_G = 3$ as suggested by counting
rules~\cite{Blankenbecler:1974tm,Farrar:yb} based on dimensional
arguments.  Thus we are left with six free parameters.

Other, more flexible, forms of the $x$ dependences at the starting
scale were also investigated. No substantial improvement in the
description of the data was observed, however the errors on the fitted
parameters as well as the correlations were increased.  Therefore we
chose to present here results obtained with a minimal number of free
parameters, as described above.

\section{Data sets}

For fitting the parameters we used 164 points of \Fg measurements
coming from $e^+e^-$ reactions, 122 proton structure function data
points from $ep$ interactions and 24 points of dijet photoproduction.

\subsection{\Fg data}

We have used all published data on the photon structure function \Fg,
coming from LEP, PETRA, and TRISTAN\footnote{The TPC2$\gamma$
data from PEP~\cite{tpc2g} have been excluded, as these data are considered to
be inconsistent with other measurements~\cite{albino}.}.

The following LEP data have been included:
\begin{itemize}
\item OPAL measurements from LEP1~\cite{opal-lep1} and from
  LEP2~\cite{opal-lep2}.  The LEP1 data are in the range
  1.86 $< Q^2 <$ 135 GeV$^2$ (32 points), while the LEP2 data are in
  the range 9 $< Q^2 <$ 780 GeV$^2$ (31 points);
\item L3 measurements~\cite{l3} in the kinematic range of
1.9 $< Q^2 <$ 120 GeV$^2$ (28 points);
\item DELPHI data~\cite{delphi} for $Q^2$ = 12 GeV$^2$ (4 points);
\item ALEPH measurements from LEP1~\cite{aleph-lep1} for
  $Q^2$ = 9.9, 20.7 and 284 GeV$^2$ (11 points), and from
  LEP2~\cite{aleph-lep2} for $Q^2$ = 17.3 and 67.2 GeV$^2$ (16 points).
\end{itemize}
 
The PETRA data are from PLUTO~\cite{pluto} at $Q^2$ = 2.4, 4.3, 9.2 and
45.0 GeV$^2$, from JADE~\cite{jade}  at $Q^2$ = 24 and 100 GeV$^2$, and
from TASSO~\cite{tasso} at $Q^2$ = 23 GeV$^2$.

Finally AMY data~\cite{amy} at $Q^2$ = 6.8, 73.0 and 390.0 GeV$^2$ and
TOPAZ data~\cite{topaz} at $Q^2$ = 5.1, 16.0 and 80.0 GeV$^2$ were used
from TRISTAN.

\subsection{$F_2^p$ data}

As stated in section~\ref{sec:glu}, we have used the Gribov factorization
relation in order to produce indirect \Fg `data' at low $x$ from the
proton structure function data $F_2^p$ measured by ZEUS~\cite{f2p}.
The $F_2^p$ data have been moved to the $Q^2$ values of the
appropriate \Fg data by using the ALLM97
parametrization~\cite{allm97}. Only data with $x < 0.01$ and $Q^2 <
100\;\GeV^2$ were used.  An additional systematic error of 3\% was
added to the measurement errors, to account for the uncertainty in the
numerical coefficient used in the derivation of the indirect \Fg
'data'.  The statistical and systematic errors were added in
quadrature.

\subsection{Dijet photoproduction data}

The dijet photoproduction measurements were taken from the ZEUS
experiment~\cite{dijets}. The cross section $d\sigma/dx_{\gamma}^{obs}$ is
measured in four bins of transverse energy $E_T$: 14-17, 17-25, 25-35,
and 35-90 GeV.  The jets are identified in the data by the $k_{\rm T}$
clustering algorithm and \ET is the transverse energy of the highest
\ET jet.

\section{Results}

The results of fits are presented for the following configurations:
\begin{itemize}
\item \Fg data, including the `indirect' data from $F_2^p$ measurements,
\item \Fg and dijet cross sections with the CTEQ5M parametrization of
the proton PDFs~\cite{CTEQ},
\item \Fg and dijet cross sections  with the ZEUS-TR parametrization
of the proton PDFs \cite{ZEUS-PDF}.
\end{itemize}
The masses of the heavy quarks used in the fit are
$m_c = 1.5\; \GeV, m_b = 4.5\; \GeV$ and $m_t = 174\;\GeV$.

The parameters are fitted to the data using MINUIT (release
96.03)~\cite{minuit} with MIGRAD and HESSE algorithms for error
calculations.

The results are summarized in Table~\ref{tbl:pars}. Also added in the
table are the values of the $\chi^2$ per degree of freedom and of the
integral~\Eq{eq:FG} representing the FG sum rule.

\def\sep{$\,\pm\,$}
\def\bsep{\qquad}
\def\mc#1&#2&{\multicolumn2c{#1}&}
\def\mcl#1&#2\cr{\multicolumn2c{#1}\cr}
\def\w{\vspace*{-10pt}}
%
%
\begin{table}[h]
\caption{Parameters of the initial distributions at $Q^2=2\;\GeV^2$, 
$\chi^2$ per degree of freedom of the respective fits
and the FG sum rule value at $Q^2 = 4\;\GeV^2$.}
\label{tbl:pars}
\medskip
\begin{tabular}{l|l@{\sep}l@{\bsep}l@{\sep}l@{\bsep}l@{\sep}l}
& \multicolumn2l{no jets} & \multicolumn2l{CTEQ5M} & \multicolumn2l{ZEUS-TR} \cr
\hline\w\cr
$A^\PL$ & 4.04 & 0.26 & 3.89 & 0.24 & 4.45 & 0.29 \cr
$B^\PL$ & 1.22 & 0.20 & 1.11 & 0.16 & 1.90 & 0.23 \cr
$A^\HL$ & 0.0645 & 0.0029 & 0.0656 & 0.0029 & 0.0647 & 0.0032 \cr
$B^\HL$ & -1.17 & 0.006 & -1.16 & 0.006 & -1.16 & 0.006 \cr
$A^\HL_G$ & 0.0173 & 0.0056 & 0.0159 & 0.0049 & 0.0271 & 0.0072 \cr
$B^\HL_G$ & -1.64 & 0.05 & -1.65 & 0.05 & -1.57 & 0.05 \cr
$\chi^2/$NDF & \mc 1.06 & 286 & \mc 1.53 & 310 & \mcl 1.63 & 310 \cr
FG SR & \mc 1.13 &  & \mc 1.14 &  & \mcl 1.04 &  \cr
\end{tabular}
\end{table} 

A general observation is that the addition of the dijet data has a
minor effect on the overall fit results and their errors. There is
though a noticeable deterioration in the $\chi^2$ value. Also the
differences due to the usage of two different proton PDFs are
relatively minor. Our preferred parametrization is the one presented
in the third column of Table~\ref{tbl:pars}, as this is the one which
fulfils best the FG sum rule. We call it the SAL parametrization and
it is presented in the following in more details.

\subsection{Comparison with \Fg data}
\begin{figure}[ht]
\centerline{%
\includegraphics*[width=\columnwidth]{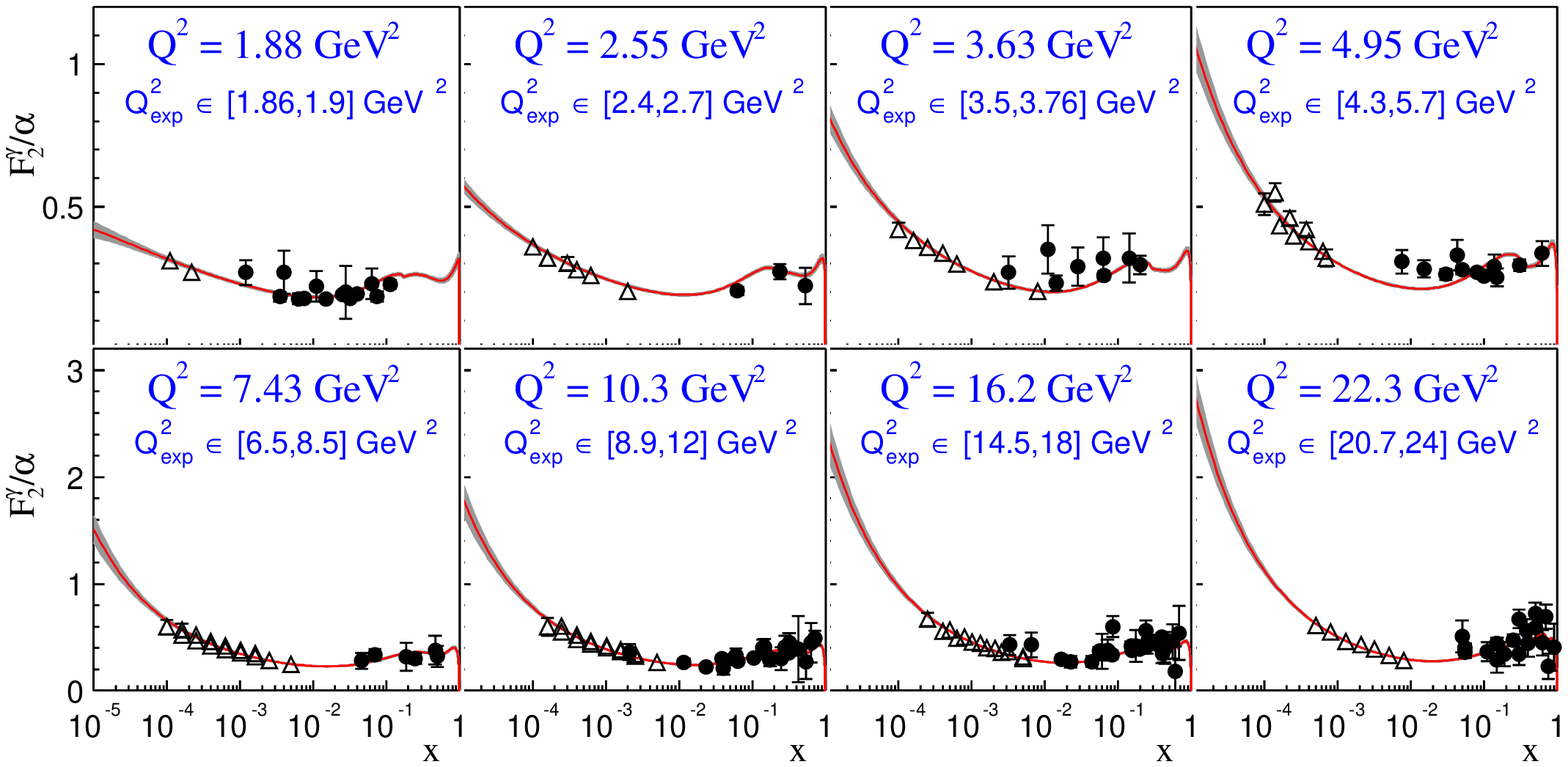}}
\vspace*{3mm}
\centerline{%
\includegraphics*[width=\columnwidth]{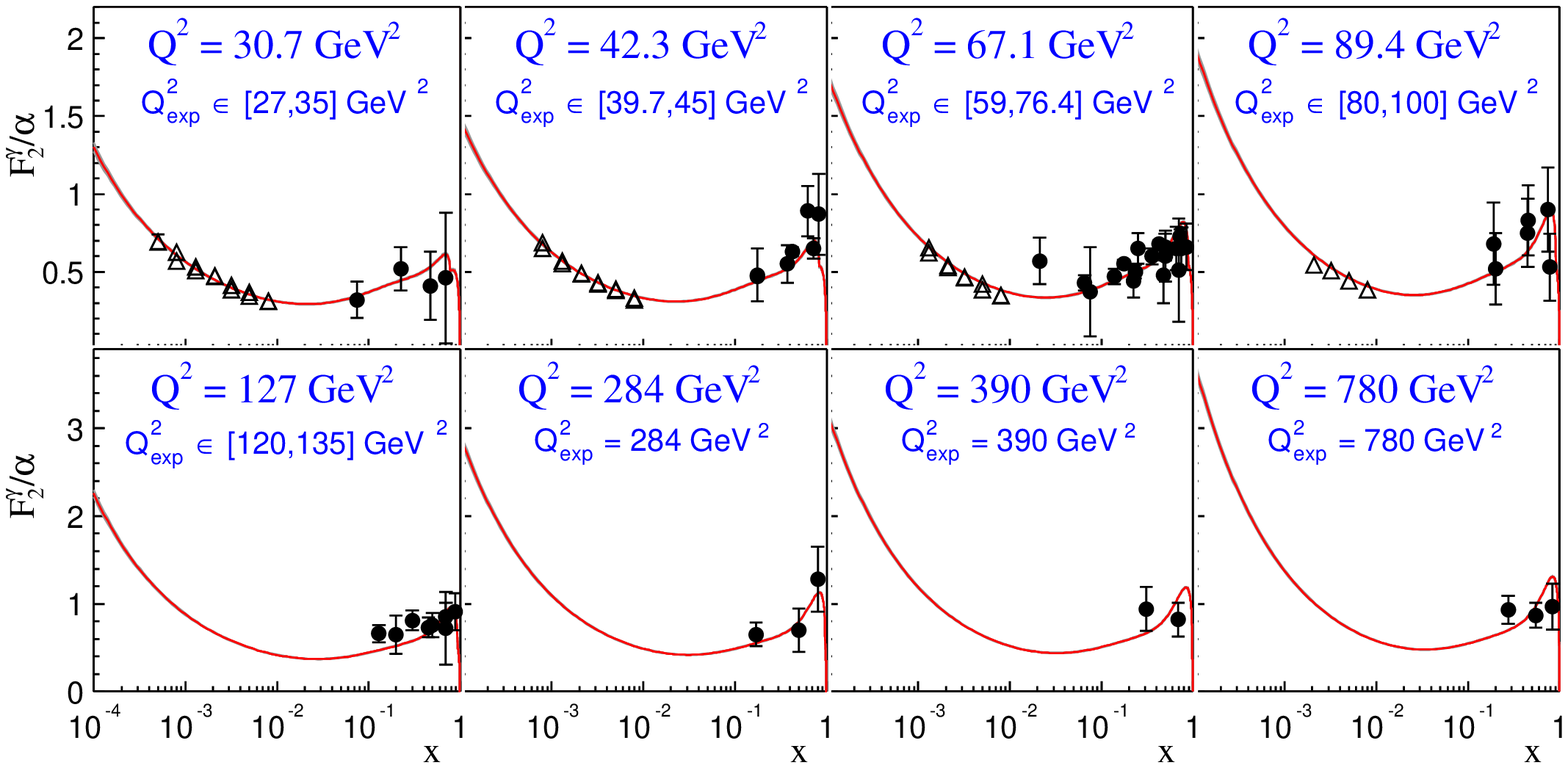}}
\caption{The SAL expectation for $\Fg(x,Q^2)$ as a function of $x$ at
selected $Q^2$ values as denoted in the figure. The plotted data (dots
for \Fg measured directly and triangles for \Fg deduced from $F_2^p$)
are from the range $Q^2_{\rm exp}$ presented in the figure. Note that
the $x$ range in the upper and lower plots is different.
}
\label{fig:F2eb10}
\end{figure}
In Figure 
\ref{fig:F2eb10} we compare the \Fg
obtained with the SAL parametrization together with the \Fg data, as a
function of $x$ in bins of $Q^2$. The real \Fg data and the ones
deduced from $F_2^p$ are shown with different symbols. To limit the
number of plots without loss of information, the data are shown within
a range of $Q^2$, while the corresponding curve is calculated for the
average $Q^2$ of that range. The shaded error band is calculated
according to the final error matrix of the fitted parameters as
returned by MINUIT.  The uncertainty becomes smaller with increasing
$Q^2$, due to the expected loss of sensitivity to the initial
conditions of the evolution.

The inclusion of the $F_2^p$ data in the fit strongly constrains the
uncertainty on the gluon distribution. As an exercise, we increased
the systematic uncertainty on the \Fg 'data' generated from $F_2^p$
from 3\% to 10\% and repeated the full fit. Small changes in the
hadronic component of the photon were observed, leading to barely
noticeable change in the PDFs in the range of $x$ and $Q^2$
investigated here. The most significant change was in the uncertainty
on the gluon content at low $Q^2$ and lowest $x$, were the error band
increased by factor two. 

\begin{figure}[ht]
\centerline{\includegraphics*[width=\columnwidth]
{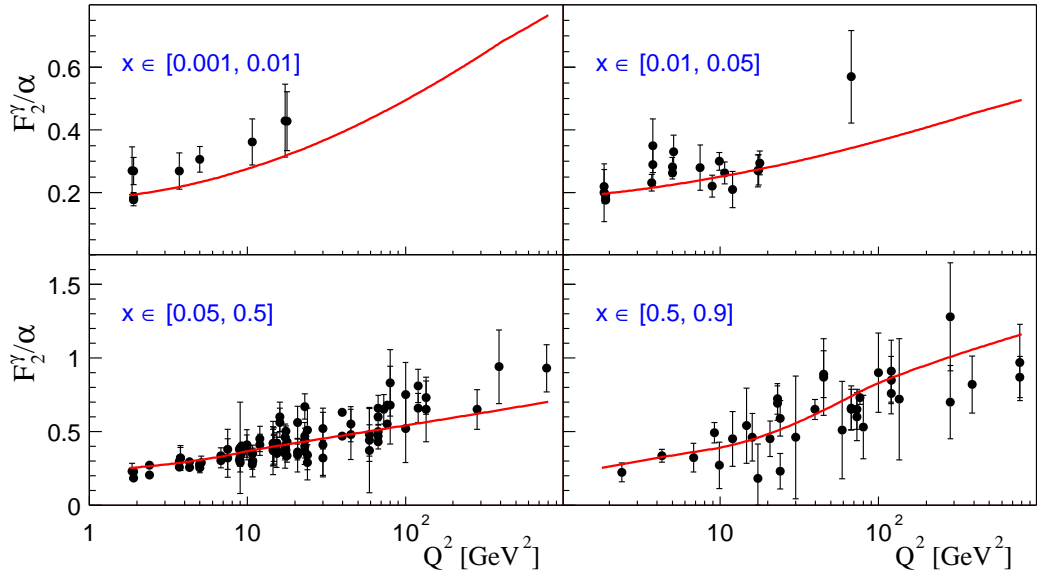}}
\caption{$Q^2$ dependence of \Fg for the $x$ ranges shown in the plots.
Black dots are \Fg data 
and the line is \Fg averaged over the depicted $x$ range.}
\label{fig:F2vsQ}
\end{figure}

The $Q^2$ dependence of \Fg is shown in \Fig{fig:F2vsQ}
for different $x$ ranges. This time only the  
\Fg data are shown. Again each plot contains data from a range of $x$
values, while the curve corresponding to the SAL parametrization is
averaged over the $x$ range in the following way:
\begin{equation}
\overline{\Fg} =
{1\over x_2 - x_1} \int\limits_{x_1}^{x_2} dx\, \Fg(x, Q^2)
\,.
\end{equation}
A good agreement between data and the results of the fit is observed.

\subsection{Comparison with dijet data }
\label{sec:dijets}

\begin{figure}[b]
\centerline{\includegraphics*[width=0.8\columnwidth]
{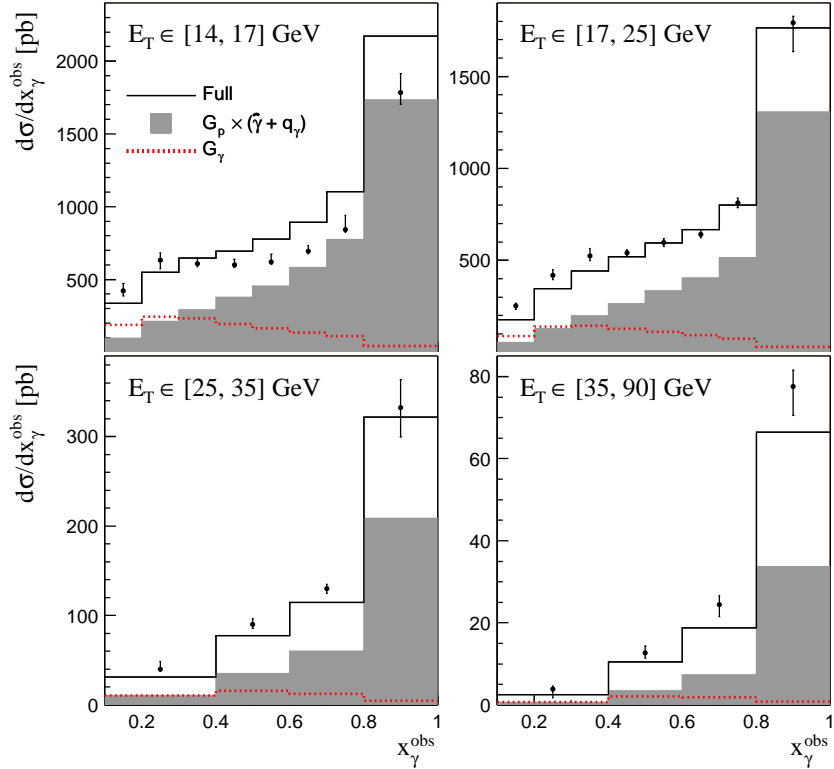}}
\caption{Dijet cross-section for ZEUS-TR proton
$\times$ SAL photon --- full line. Other
contributions:
$G_p \times (\hat\gamma + q_\gamma)$  --- shaded;
$(q_p + G_p) \times \,G_\gamma$  --- dotted line.}
\label{fig:dj1Z}
\end{figure}
The dijet photoproduction cross sections are compared to calculations
obtained using the
Frixione-Ridolfi code with jets identified by the $k_{\rm T}$-clustering
algorithm, the same as used by the ZEUS collaboration to analyse the data.
The predictions for the SAL photon parametrization and 
the ZEUS-TR proton PDFs are presented in \Fig{fig:dj1Z}.
Also shown are the individual contributions of
partonic cross sections induced by gluons in the photon $G_\gamma$ and
gluons in the proton $G_p$. 

With the exception of the lowest $E_T$ bin, the calculations tend to
underestimate the measured cross sections, especially at $x <0.4$.
It should be noted, however, that within the  
uncertainty\footnote{Changing the factorization/renormalization
scale in the range between $0.5\ET^2$ and $2\ET^2$
may change the cross sections by about 20\%~\cite{dijets}.}
of the NLO calculation there is no clear evidence for disagreement.
Nevertheless, in this study we aimed at improving the agreement by 
adjusting the gluon content of the photon.
However, only in the range $0.1<x<0.2$ the gluons from the photon
contribute roughly half the cross section. In all other bins the
contribution of the gluons in the proton dominates.
The dijet data are therefore more sensitive to the gluons in the proton
than to the gluons in the photon. It would therefore be tempting to use
these data in the global fits of parton densities of the proton. However
the present uncertainty on the NLO calculations of the dijet cross
section is too large to reliably extract the gluon density in the proton.

\subsection{Parton distribution functions}

\def\cmpFig#1,#2;{%
\begin{figure}[t]
\centerline{\includegraphics*[width=0.8\columnwidth]{axpd#1_#2.eps}}%
\vspace*{1mm}
\caption{Comparison of SAL to other NLO parametrizations at $Q^2 = #2\;\GeV^2$.}
\label{fig:cmp#2}
\end{figure}}

\begin{figure}[b]
\centerline{\includegraphics*[width=0.8\columnwidth]{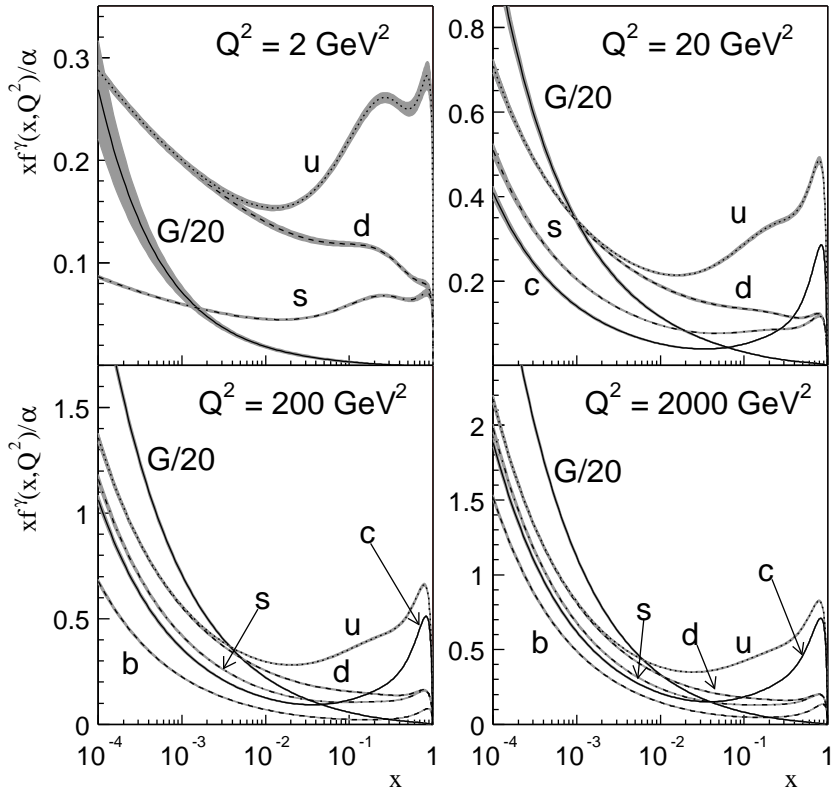}}
\caption{Parton distribution functions in the photon 
for different values of $Q^2$ as denoted in the figure.}
\label{fig:PDFs}
\end{figure}

The SAL parton distribution functions in the photon are shown in \Fig{fig:PDFs}.
The features to be noted are the behaviour of quarks at large $x$,
typical of the point-like contribution of the photon, and the
dominance of the gluon distribution at low $x$.

A comparison between the SAL PDFs and the other available NLO \DISg photon
parametrizations,
GRV \cite{GRV92}, GRS\footnote{This parametrization uses Fixed Flavour
Number Scheme (FFNS), where only $u,d$ and $s$ PDFs exist.}
\cite{GRS99}, and CJK \cite{CJK}, is shown in
\Fig{fig:cmp2.5} for $Q^2=2.5\;\GeV^2$ and in \Fig{fig:cmp100}
for $Q^2=100\;\GeV^2$. 

\cmpFig1,2.5;
At low $Q^2$ there are big differences between
the various PDFs%
\footnote{A non-vanishing $b$-quark density at $Q^2=2.5\;\GeV^2$
is a feature of the CJK parametrization.}. 
They are especially pronounced for $x<10^{-3}$,
where no \Fg data are available and the result is subject to
additional theoretical assumptions. The SAL parametrization has the
lowest gluon distribution down to $x\sim 10^{-4}$,  below which value we
observe a steep rise, steeper than in the other PDFs.
\cmpFig1,100;
At high $Q^2$, where the sensitivity to initial conditions is
diminished, there are still noticeable differences.

To further compare the various parametrizations, the momentum fraction
carried by the gluons and the light quarks are summarized in
Table~\ref{tab:frac}. They can be used to assess the fulfilment of the
FG sum rule. 
According to the FG sum rule, the value of the sum G+\Sigl\ should be
$\approx 0.9$ at $Q^2 = 2.5\;\GeV^2$ and $\approx 1$ at $Q^2 = 4\;
\GeV^2$. The SAL parametrization is found to be closest to these
values. 

\begin{table}[h]
\centering
\caption{Parton momentum fractions at $Q^2 = 2.5\;\GeV^2$ and $4\;\GeV^2$. 
\Sigl $= 2 (p_u+p_d+p_s)$ for the NLO PDFs. }
\label{tab:frac}
\medskip

\begin{tabular}{c|*4r|*4r}
 & \multicolumn4{c|}{$Q^2 = 2.5\; \GeV^2$} & \multicolumn4c{$Q^2 = 4\;
 \GeV^2$}\cr
\hline
  & GRV  & GRS  & CJK  & SAL  & GRV  & GRS  & CJK  & SAL \cr
G & 0.47 & 0.29 & 0.87 & 0.06 & 0.50 & 0.32 & 0.90 & 0.12 \cr
d & 0.13 & 0.07 & 0.14 & 0.10 & 0.13 & 0.08 & 0.15 & 0.10 \cr
u & 0.25 & 0.24 & 0.21 & 0.26 & 0.28 & 0.27 & 0.24 & 0.28 \cr
s & 0.06 & 0.04 & 0.04 & 0.07 & 0.06 & 0.05 & 0.05 & 0.07 \cr
\hline
\Sigl
  & 0.88 & 0.70 & 0.77 & 0.85 & 0.96 & 0.78 & 0.86 & 0.92 \cr
G+\Sigl
  & 1.34 & 0.99 & 1.64 & 0.91 & 1.46 & 1.10 & 1.76 & 1.04 \cr
\end{tabular}
\end{table}

\section{Conclusions} 

The \Fg measurements supplemented by the low $x$ expectations for \Fg
based on the measurements of $F_2^p$ and Gribov factorization, and by
dijet photoproduction data have been used to extract a new NLO
parametrization of the parton distributions in the photon. 

A good description of the data is obtained with the new
parametrization, with the exception of the dijet data which turned out
to be mostly sensitive to the gluon distribution in the proton. The
obtained parton distributions
fulfil the Frankfurt-Gurvich sum rule which was not imposed in the
fit.

\section*{Acknowledgements}
\noindent This work was partially supported by the Israel Science 
Foundation (ISF). One of us (W.S.) greatly acknowledges discussions with
Micha{\l} Prasza{\l}owicz and Krzysztof Golec-Biernat.


\begin{thebibliography}{99}
\bibitem{reviewf2} See \eg R. Devenish and A. Cooper-Sarkar, {\em Deep
    Inelastic Scattering}, Oxford University Press (2004).
\bibitem{hawar} See e.g. M. Krawczyk, A. Zembrzuski and M. Staszel,
  \prep{345}{2001}{265}; 
\  A. De Roeck, \epj{C33}{2004}{S394}.
\bibitem{GRV92} M. Gluck, E. Reya, A. Vogt, 
\prev {D45} {1992} {3986}, \prev {D46} {1992} {1973}.
\bibitem{AFG} P. Aurenche, J.-Ph. Guillet, M. Fontannaz,
\zp {C64} {1994} {621}.
\bibitem{GS96}
L.E.~Gordon and J.K.~Storrow,
Nucl.\ Phys.\ B {\bf 489} (1997) 405.
\bibitem{GRS99}
M.~Gluck, E.~Reya and I.~Schienbein,
Phys.\ Rev.\ D {\bf 60} (1999) 054019;
Erratum-ibid.\ D {\bf 62} (2000) 019902.
\bibitem{CJK}
F.~Cornet, P.~Jankowski and M.~Krawczyk,
Phys.\ Rev.\ D {\bf 70} (2004) 093004.
\bibitem{al-gribov} A. Levy, \pl{B404}{1997}{369}.
\bibitem{f2p}
ZEUS\ Collaboration, J. Breitweg {\it et al.},
  \epj{C7}{1999}{609};
\ \epj{C21}{2001}{443}.
\bibitem{gribov-fact} V.N. Gribov, L.Ya. Pomeranchuk, \prl {8} {1962} {343}.
\bibitem{dijets}ZEUS Collaboration, S. Chekanov {\it et al.},
\epj{C23}{2002}{615}.
\bibitem{dl} A. Donnachie and P. Landshoff, \pl{B296}{1992}{227}.
\bibitem{FP} W. Furmanski, R. Petronzio, \zp {C11} {1982} {293}.
\bibitem{Frix} 
S. Frixione, Z. Kunszt and A. Signer, \np{B467}{1996}{399};
\\
S. Frixione, \np{B507}{1997}{295}.
\bibitem{witten}
E.~Witten, \np{B120}{1977}{189}.
\bibitem{bardeen}
W.A.~Bardeen and A.J.~Buras, \prev{D20}{1979}{166}
[Erratum-\prev{D21}{1980}{2041}].
\bibitem{dewitt}
R.J.~DeWitt, L.~M.~Jones, J.D.~Sullivan, D.E.~Willen and H.W.~Wyld,
\prev{D19}{1979}{2046}
[Erratum-\prev{D20}{1979}{1751}].
\bibitem{CFP} 
G. Curci, W. Furmanski, R. Petronzio, \np{B175}{1980}{27};
\\
W. Furmanski, R. Petronzio, \pl {97B} {1980} {437}.
\bibitem{ACOT}
M.A.G.~Aivazis, J.C.~Collins, F.I.~Olness, and W.-K.~Tung,
\prev{D50}{1994}{3102}.
%
\bibitem{Collins}
J.C. Collins, \prev{D58}{1998}{094002}.
%
\bibitem{TR}
R.S.~Thorne and R.G.~Roberts,
\prev{D57}{1998}{6871}.
%
\bibitem{Tung2002}
W.-K.~Tung, S.~Kretzer and C.~Schmidt, \jpg{28}{2002}{983}.
%
\bibitem{Budnev}
V.M. Budnev, I.F. Ginzburg, G.V. Meledin and V.G. Serbo,
\prep {15}{1974}{181}.
%
\bibitem{Thorne-DIS05}
For a recent discussion see 
R. Thorne's talk at the DIS05 workshop,
hep-ph/0506251.
\bibitem{SaS}
G.A. Schuler, T. Sjöstrand, \zp{C68}{1995}{607}.
\bibitem{fg-sumrule} L.L. Frankfurt, E.G. Gurvich, 
hep-ph/9505406;
\ \pl{B386}{1996}{379};
\jpg{22}{1996}{903}.
\bibitem{ws-evol} W. Slominski, to be published.
\bibitem{Blankenbecler:1974tm}
R.~Blankenbecler and S.J.~Brodsky,
\prev{D10}{1974}{2973}.
\bibitem{Farrar:yb}
G.R.~Farrar and D.R.~Jackson,
\prl{35}{1975}{1416}.

\bibitem{tpc2g} H. Aihara et al., \prl{58}{1987}{97}; 
\ H. Aihara et al., \zp{C34}{1987}{1}.
\bibitem{albino}S. Albino, M. Klasen and S. Söldner-Rembold,
  \prl{89}{2002}{122004}.  
\bibitem{opal-lep1}OPAL\ Collaboration,
  \zp{C61}{1994}{199};\ \pl{B412}{1997}{412};\ \zp{C74}{1997}{33};
  \ \epj{C18}{2000}{15}.  
\bibitem{opal-lep2}OPAL\ Collaboration,
  \pl{B411}{1997}{387};\ \epj{C18}{2000}{15}; \ \pl{B533}{2002}{207}.
\bibitem{l3}L3\ Collaboration, \pl{B436}{1998}{404};
  \ \pl{B447}{1999}{147}; \ \pl{B483}{2000}{373}.
\bibitem{delphi}DELPHI\ Collaboration, \zp{C69}{1996}{223}.
\bibitem{aleph-lep1}ALEPH\ Collaboration, \pl{B458}{1999}{152}.
\bibitem{aleph-lep2}ALEPH\ Collaboration, \epj{C30}{2003}{145}.
\bibitem{pluto}PLUTO\ Collaboration, \pl{B142}{1984}{111};
\ \np{B281}{1987}{365}. 
\bibitem{jade}JADE\ Collaboration, \pl{B121}{1983}{203};\ \zp{C24}{1984}{231}.
\bibitem{tasso}TASSO\ Collaboration, \zp{C31}{1986}{527}. 
\bibitem{amy}AMY\ Collaboration, \pl{B346}{1995}{208};\ \pl{B400}{1997}{395}.
\bibitem{topaz}TOPAZ\ Collaboration, \pl{B322}{1994}{447}. 
\bibitem{allm97}H. Abramowicz and A.Levy, {\it The ALLM parameterization
of $\sigma_{tot}(\gamma*p)$: An update.}, 
Preprint DESY-97-251 [hep-ph/9712415], DESY, 1997.
\bibitem{CTEQ}
CTEQ Collaboration, H.L.~Lai {\it et al.},
Eur.\ Phys.\ J.\ C {\bf 12} (2000) 375.
\bibitem{ZEUS-PDF}
ZEUS Collaboration, S.~Chekanov {\it et al.},
Phys.\ Rev.\ D {\bf 67} (2003) 012007.
\bibitem{minuit} For the fits the MINUIT package is used: MINUIT (release
96.03): F.~James
and M.~Roos,\cpc{10}{1975}{343}.

\end{thebibliography}
\end{document}